# Detection Limit for Optically Sensing Specific Protein Interactions in Free-solution

Harish Sasikumar[†], Manoj M. Varma[*,†,‡,§]

[†]Department of Electrical Communication Engineering, Indian Institute of Science, Bangalore, 560012, India

[‡] Center for Nano Science and Engineering, Indian Institute of Science, Bangalore, 560012, India

[§] Robert Bosch Center for Cyber Physical Systems, Indian Institute of Science, Bangalore, 560012, India

**ABSTRACT:** Optical molecular sensing techniques are often limited by the refractive index change associated with the probed interactions. In this work, we present a closed form analytical model to estimate the magnitude of optical refractive index change arising from protein-protein interactions. The model, based on the Maxwell Garnett effective medium theory and first order chemical kinetics serves as a general framework for estimating the detection limits of optical sensing of molecular interactions. The model is applicable to situations where one interacting species is immobilized to a surface, as commonly done, or to emerging techniques such as Back-Scattering Interferometry (BSI) where both interacting species are un-tethered. Our findings from this model point to the strong role of as yet unidentified factors in the origin of the BSI signal resulting in significant deviation from linear optical response.

## INTRODUCTION

Optical refractometry, i.e. the measurement of optical refractive index (RI), is widely used for material characterization, for instance, to quantify the purity of a sample[1]. Highly sensitive measurement of refractive index would allow the measurement of small changes associated with the adsorption of molecules onto functionalized surfaces which is very useful for developing optical biosensing techniques. Indeed, several optical molecular sensing techniques described in literature such as Surface Plasmon Resonance (SPR)[2], Optical Waveguide Lightmode Spectroscopy (OWLS)[3], Dual-Polarization Interferometry (DPI)[4], Silicon micro-ring resonators (MRR)[5] etc. are based on RI changes that occur due to the binding of target proteins or DNA to surfaces functionalized with receptors specific to the target. Typical limits-of-detection (LoD) of these techniques range from $10^{-7} – 10^{-5}$ Refractive Index Units (RIU). As mentioned above, almost all of the published optical molecular sensing techniques are based on solid phase reactions between the target and its receptor, i.e. the receptor molecules are immobilized on a solid surface in a thin layer (~ few nanometers) thus "functionalizing" the surface which captures target molecules from, typically, liquid samples. The binding of the target molecules to the functionalization layer increases the optical density, which is a combined effect of increase in surface density of molecules as well as increase in the average layer thickness. This change in optical density can be measured using a wide variety of techniques mentioned earlier.

Although surface immobilized sensors dominate the available suite of optical molecular sensing techniques, they do suffer from a few problems. A significant issue with surface immobilized sensors is that the receptors, which are often proteins, are not in their native conformational state. The conformational state strongly influences the functionality of protein receptors and distortions in the native conformational state lead to inaccuracies in the estimation of kinetic parameters such as affinity constants[6]. In the context of diagnostic sensors, ensuring satisfactory performance from surface immobilized sensors often requires multi-step, site-specific immobilization methods to ensure control of receptor orientation and functionality on the sensor surface increasing the complexity of the sensing method[7]. Therefore, there is a strong case for the development of optical techniques that can measure interactions between target and receptor proteins in free solution, i.e. free of surface immobilization. To address this need, Bornhop et al. demonstrated a novel technique, which they referred to as Back-Scattering Interferometry (BSI), to probe molecular interactions in free solution[6b]. They demonstrated that BSI can distinguish specific interactions (i.e. between the target and its specific receptor) from non-specific interactions. The working principle of BSI is that laser illumination of a capillary filled with a liquid sample produces a set of interference fringes due to multiple reflections from various interfaces in the system. A shift in the refractive index of the liquid sample, such as due to the molecular interactions taking place in the liquid, results in a spatial shift of the fringe pattern which is detected using image analysis of the fringe pattern captured using a CCD camera. This technique is capable of attaining high sensitivities (better than $10^{-6}$ RIU, or ppm level shift detection) permitting time-resolved measurement of molecular interactions occurring in the solution from which kinetic parameters of the interaction can be extracted[8].

Although detection of specific molecular interactions using BSI has been described in several papers, a model relating the molecular interactions to optical refractive index changes has been lacking[9]. A recent paper[10] explained the origin of BSI signal by fitting it to a linear function of four variables, namely, the average values and change upon binding of the solvent-addressable surface area and the radius of gyration of the interacting proteins. In addition to these four parameters, a con-stant term was also included, which accounted for most of the numerical agreement with the fit and experimental data[11]. This model therefore did not present a first principles description of signal generation based on a solid theoretical framework. In this paper, we present a first principle analysis of RI changes due to protein interactions in free-solution. This analysis focuses on estimating magnitude of RI changes expected due to protein interactions by considering changes in conformation and dielectric properties (due to changes in hydration state and so on). A Maxwell-Garnett effective medium model is used to convert changes in conformation and dielectric properties due to protein interactions into RI changes. The model presented here is applicable not only to BSI but also to any technique attempting to probe molecular interactions, not limited to proteins, in free solution. Calculations using this model revealed that the theoretically expected RI changes associated with previously reported BSI experiments are about 2-3 orders of magnitude smaller than the reported limit of detection of BSI. This result suggests the search for alternate models to explain BSI signal origin. The model also reveals why surface immobilization has served as a successful strategy for optical molecular sensing by increasing the local density of interacting molecules.

## EFFECTIVE REFRACTIVE INDEX OF A PROTEIN SOLUTION

We consider a protein solution (shown in Figure 1) to be composed of a number of dielectric spheres with relative dielectric permittivity $\epsilon_R$, representing the protein molecules, in a medium with relative dielectric permittivity of $\epsilon_m$. The medium is typically air or water and the index "R" refers to the receptor. This dielectric composite consists of protein molecules occupying small fraction (<0.002%, as discussed further in this section) of the total volume. Moreover, on an average, the solution remains uniform and isotropic. Hence, the effective permittivity, $\epsilon_{eff}$ of the solution, based on Maxwell Garnett effective medium theory[12], can be written as,

$$\frac{\epsilon_{eff} - \epsilon_m}{\epsilon_{eff} + 2\epsilon_m} = f \frac{\epsilon_R - \epsilon_m}{\epsilon_R + 2\epsilon_m} \quad (1)$$

Here, f is the volume fraction of the spheres.

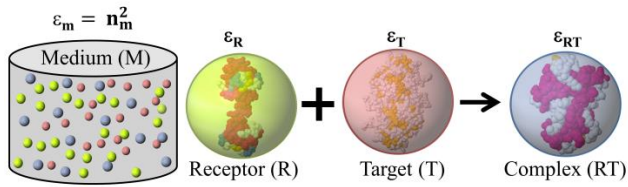

Figure 1. A representative depiction of receptor, target and receptor-target complex in a fluid medium. The protein structures were taken from[13].

The volume fraction occupied by the receptor molecules is given by,

$$f_R = c_R N_{Avo} \upsilon_R \quad (2)$$

where $c_R$ and $\upsilon_R$ are the molar concentration of the receptors in solution and the molecular volume respectively and $N_{Avo}$ is the Avogadro's number.

The typical molecular volume of proteins probed in BSI studies[10] (and other molecular sensing methods) is about 33 nm$^3$. For a receptor concentration as large as a micro-molar, the volume fraction turns out to be about $2 \times 10^{-5}$. In other words, the volume fractions of protein samples used in BSI or in other molecular sensing techniques are of the order of 10$^{-5}$ or smaller. In the regime of $f \ll 1$ and noting that refractive index is the square root of the permittivity, from Eq. (1), one can write the effective refractive index of the solution, $n_{eff}$, as

$$n_{eff} = n_m \left(1 + \frac{3}{2} f_R \left(\frac{\epsilon_R - \epsilon_m}{\epsilon_R + 2\epsilon_m}\right)\right) \quad (3)$$

where $n_m$ is the refractive index of the medium.

The concentration of receptors, $c_R$, can be written in terms of the mass density of receptors in the solution, $\mu_R$ and the molecular mass $M_R$ as $c_R = \mu_R/M_R$ which allows us to rewrite Eq. (2) as,

$$f_R = \frac{\mu_R}{\rho_R} \quad (4)$$

where $\rho_R$ is the density of the receptor molecule.

We can define a quantity "refractive index increment", $\gamma = \frac{\partial n_{eff}}{\partial \mu_R}$ which, from Eq. (3) and (4) can be written as,

$$\gamma = \frac{3}{2} \frac{n_m}{\rho_R} \left(\frac{\epsilon_R - \epsilon_m}{\epsilon_R + 2\epsilon_m}\right) \quad (5)$$

The quantity $\gamma$ has been measured for several proteins and is found to be tightly clustered around the value of about 0.19 ml/g[14]. The universality of this quantity, in the light of Eq. (5), must lie in the tight clustering of the density and dielectric permittivity of proteins. Indeed the density of proteins is tightly clustered around 1.35 g/cc[15]. Using these numbers in Eq. (5) yields the relative dielectric permittivity of proteins to be tightly clustered around 2.6, or in terms of refractive index, around 1.6. In other words, we can consider any protein solution to be composed of dielectric spheres with a relative permittivity of about 2.6 and molecular volume corresponding to the protein molecule being considered.

## REFRACTIVE INDEX OF MIXTURE CONTAINING INTERACTING PROTEINS

Now, let us consider what happens when a solution containing target molecules with concentration $c_T$ are introduced into this solution. There will be an interaction between the target and receptor molecules leading to the formation of the target-receptor complex. As shown in Figure 1, there will be three species in the medium, namely the free receptor molecules (denoted by subscript 'R'), free target molecules (denoted by subscript 'T') and bound complex (denoted by subscript 'RT'). When there are multiple species involved, each with relative permittivity $\epsilon_i$ and volume fraction $f_i$, the effective medium formula of Eq. (1) is extended as

$$\frac{\epsilon_{eff} - \epsilon_m}{\epsilon_{eff} + 2\epsilon_m} = \sum_i f_i \frac{\epsilon_i - \epsilon_m}{\epsilon_i + 2\epsilon_m} \quad (6)$$

Accordingly, Eq. (3), gets modified as,

$$n_{eff} = n_m \left(1 + \frac{3}{2}\left\{f_R\left(\frac{\varepsilon_R - \varepsilon_m}{\varepsilon_R + 2\varepsilon_m}\right) + f_T\left(\frac{\varepsilon_T - \varepsilon_m}{\varepsilon_T + 2\varepsilon_m}\right) \right.\right.$$
$$\left.\left. + f_{RT}\left(\frac{\varepsilon_{RT} - \varepsilon_m}{\varepsilon_{RT} + 2\varepsilon_m}\right)\right\}\right) \quad (7)$$

Where $f_i$ and $\varepsilon_i$ are the volume fraction and relative dielectric permittivity of the respective molecular species and $n_m$ is the refractive index of the medium.

The principles of mass action kinetics can be applied to the monovalent receptor, target and bound complexes[16]. Accordingly, the concentration of free receptors, free targets and bound complexes will evolve according to Eq. (8), eventually reaching equilibrium and so will the respective volume fractions according to Eq.(2).

$$\frac{dc_{RT}}{dt} = k_{on}(R_0 - c_{RT})(T_0 - c_{RT}) - k_{off}c_{RT} \quad (8)$$

Here, $c_{RT}$ is the concentration of the bound target-receptor complex, $R_0$ and $T_0$ are the initial receptor and target concentration. $k_{on}$ and $k_{off}$ are the association and dissociation rates of the reaction, respectively. At any point in time, the free receptor and free target molecules will be given by,

$$c_R = R_0 - c_{RT} \quad (9a)$$
$$c_T = T_0 - c_{RT} \quad (9b)$$

The temporal profile of the concentration change due to protein interactions can be obtained by solving Eq. (8), which yields,

$$c_{RT}(t) = \frac{R_0 + T_0 + K_D}{2} - \frac{D}{2}\tanh\left(\frac{k_{on}D}{2}t + \tanh^{-1}\left(\frac{R_0 + T_0 + K_D}{D}\right)\right) \quad (10)$$

Where $D = \sqrt{(R_0 - T_0)^2 + 2K_D(R_0 + T_0) + K_D^2}$; $K_D$ is the ratio of the off-rate to on-rate in Eq. (8), referred to as the dissociation constant of the reaction.

The equilibrium concentration, $C_{max}$, is the steady state solution to Eq. (8). A detailed derivation is provided in the Supplementary Information (SI) text. $C_{max}$ is given as

$$C_{max} = \frac{R_0 + T_0 + K_D - D}{2} \quad (11)$$

Using Eqns. (2), (9a) and (9b) in Eq. (7), one obtains the change in refractive index due to receptor-target interactions as,

$$\delta n = \frac{3}{2}n_m N_{Avo} c_{RT}\left\{\upsilon_{RT}\left(\frac{\varepsilon_{RT} - \varepsilon_m}{\varepsilon_{RT} + 2\varepsilon_m}\right) - (\upsilon_R + \upsilon_T)\left(\frac{\varepsilon_P - \varepsilon_m}{\varepsilon_P + 2\varepsilon_m}\right)\right\} \quad (12)$$

In Eq.(12), we have used

$$\varepsilon_P = \varepsilon_R = \varepsilon_T = 2.6 \quad (13)$$

the universal value of relative dielectric permittivity applicable to any protein molecule discussed in the previous section. There are a few points to note in Eq. (12)

1) Eqn. (12) represents a change in RI with respect to the baseline. The baseline is considered to be the time at which the target and receptor solutions are mixed. To compare Eqn. (12) with experiments, the baseline (or the RI value at the time of mixing (t = 0)) should be subtracted out.

2) One of the major issues raised by[10] was that solvation effects and conformational change are not included in the refractive index calculations mentioned in the references they used. However, as we can see from Eq. (12), the result of interaction between receptor and target can manifest as a change in the relative permittivity ($\varepsilon_{RT} \neq \varepsilon_P$), and/or a change in molecular volume ($\upsilon_{RT} \neq \upsilon_R + \upsilon_T$). Both of these changes may arise due to changes in the solvation state of the complex, relative to the unbound state, and concomitant conformational changes, as extensively discussed in[10]. When $\varepsilon_{RT} = \varepsilon_P$ and $\upsilon_{RT} = \upsilon_R + \upsilon_T$; a situation we may expect when there are absolutely no interactions between the receptor and the target, we see that Eq. (12) predicts zero change in the RI. Thus, we see that a proper treatment of the receptor-target interaction in the framework of Maxwell-Garnett effective medium theory indeed reveals, as they have argued, that solvation effects (through $\varepsilon_{RT}$) and conformational changes (predominantly through $\upsilon_{RT}$ and to some extent through $\varepsilon_{RT}$) are the underlying basis for the detection of specific protein interactions in free solution.

3) The RI change is directly proportional to the concentration of the receptor-target complex. Therefore, the RI change will follow the temporal profile of receptor-target binding and will show the saturation behaviour expected in reversible reactions, which is seen in experiments of[6b].

In order to calculate the temporal profile of the RI change due to protein interactions, we plug in Eq.(10) in Eq. (12). However, there are two unknown factors in Eq. (12), namely the relative difference between $\varepsilon_{RT}$ and $\varepsilon_P$ and between $\upsilon_{RT}$ and $\upsilon_R + \upsilon_T$. We can write,

$$\varepsilon_{RT} = \alpha\varepsilon_P, \text{and} \quad (14a)$$
$$\upsilon_{RT} = \beta(\upsilon_R + \upsilon_T) \quad (14b)$$

The factors $\alpha$ and $\beta$ represent the contributions from solvation effects and conformational changes, respectively, during protein-protein interactions. The $\alpha$ factor could also have a dependence on conformational changes.

Using Eq. (14), Eq. (12) can be re-written as

$$\delta n = Fc_{RT}(v_R + v_T)\Delta \quad (15)$$

Where F is given as $F = 1.5\, n_m N_{Avo}$ and $\Delta$ is the factor capturing the effect of dielectric and volumetric changes in the molecules. Using Eq. (12) and Eq. 14, $\Delta$ is,

$$\Delta = \left(\beta\frac{\alpha\varepsilon_P - \varepsilon_m}{\alpha\varepsilon_P + 2\varepsilon_m} - \frac{\varepsilon_P - \varepsilon_m}{\varepsilon_P + 2\varepsilon_m}\right) \quad (16)$$

Similarly, the equilibrium value (also the maximum value at reaction saturation), $\delta n_{max}$ can be obtained by substituting Eqn. (11) in Eqn. (12).

$$\delta n_{MAX} = FC_{max}(v_R + v_T)\Delta \quad (17)$$

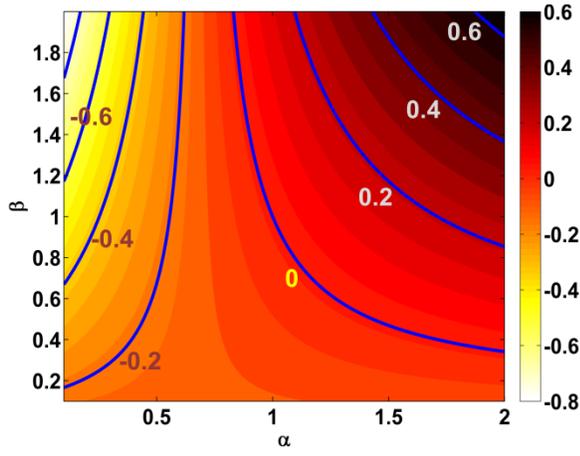

Figure 2. Variation of Δ, as described in Eq. (16) (for $\varepsilon_P = 2.6$ and $\varepsilon_m = 1.78$) as a function of α and β within their expected ranges.

Figure 2 shows the variation of Δ as a function of α and β over a limited range of parameters around their expected values based on available data. The value of β, which represents the relative change in size due to protein interaction, can be estimated using the data presented in[10]. The value of β from this data is around 0.8-1.2, representing a reduction or increase in the size of the complex relative to the unbound protein pair.

There is no reported data allowing the estimation of the value of α. However the factor involving $\varepsilon_{RT}$ in Eqn. (12) reaches a limiting value for arbitrarily large values of $\varepsilon_{RT}$. This is evident from Eqn. 16, where the effect of α saturates in both the upper ($\alpha \to \infty$) and lower ($\alpha \to 0$) limits (See SI). In other words, arbitrarily large changes in dielectric permittivity due to protein interactions will not lead to large changes in the bulk refractive index of the solution. Instead, the RI change saturates to a limiting value as $\varepsilon_{RT}$ is increased. Arbitrarily large values of $\varepsilon_{RT}$ provide us with an upper bound for the expected RI change. With these observations, we are finally in a position to calculate the RI changes expected in typical biosensing applications.

## RI CHANGES DUE TO PROTEIN INTERACTIONS

Now, we are in a position to use Eqn. (15) to estimate the refractive index changes expected in protein-protein interactions. In order to match the simulations as close to experimental observations as possible, we chose the concentrations and reaction rate constants from previously published BSI data[6b]. Specifically, we picked the interactions of Calmodulin (CaM) with $Ca^{2+}$, Trifluoperazine Dihydrochloride (TFP), M13 peptide from skeletal muscle myosin light chain kinase (sk-MLCK) and the interaction of interleukin 2 (IL2) with its antibody (anti-IL2).

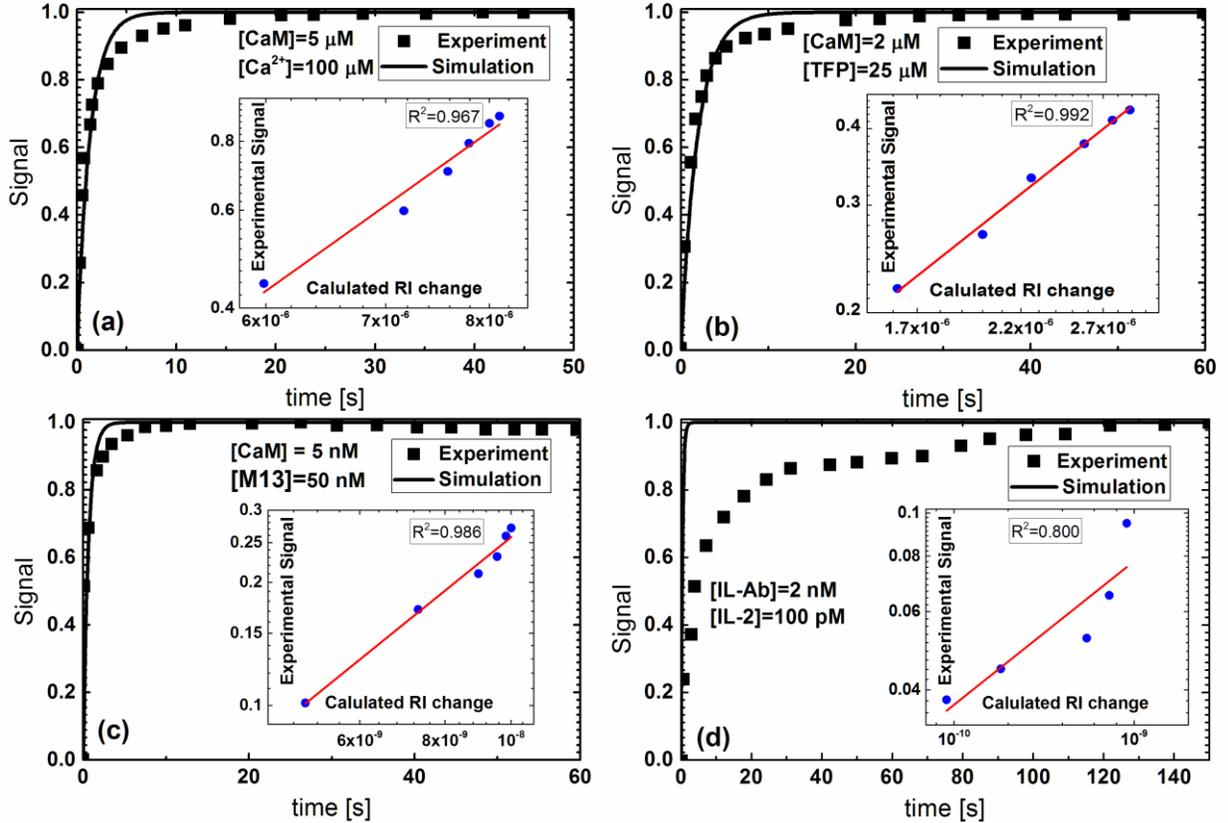

Figure 3. Comparison of the normalized simulation (based on Eq. (15)) and experimental plots (extracted from previously published BSI data[6b]) of the temporal evolution of RI change due to protein interaction; a) CaM-$Ca^{2+}$ b) CaM-TFP c) CaM-M13 and d) IL2Ab – IL2. Figures in the inset shows the comparison of refractive index variation predicted by the simulation and the signal observed in the BSI Data. The predicted RI change is orders of magnitude lower than BSI detection limit.

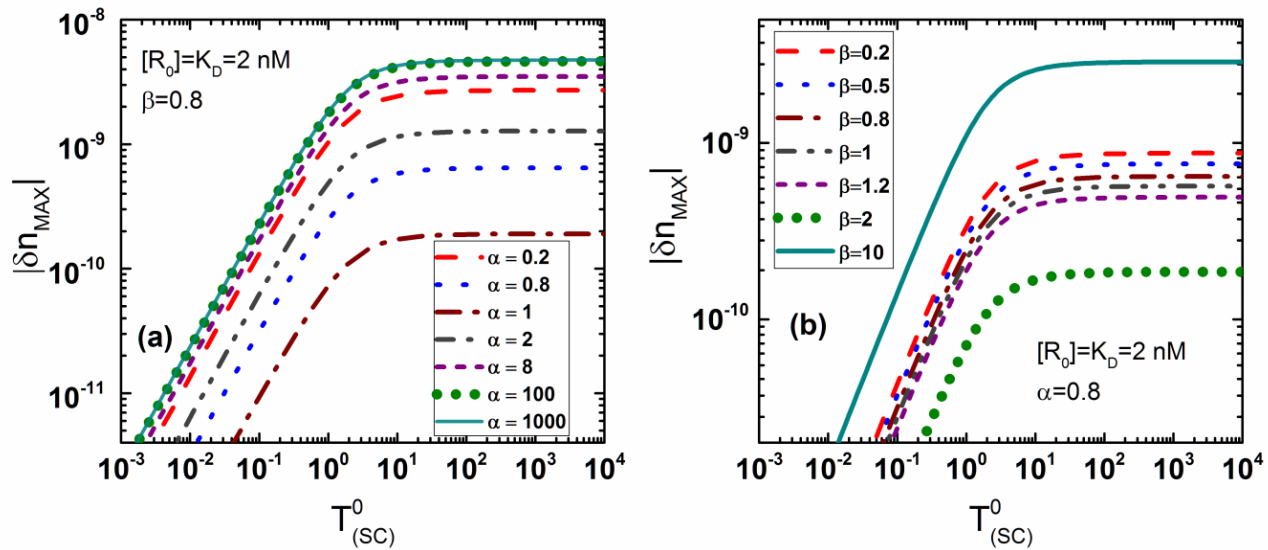

Figure 4. (a) Variations of typical $|\delta n_{MAX}|$ with $\alpha$. The constant terms used in conceiving the graph, $\beta = 0.8$ was taken as a reasonable value based on previously published BSI data[6b]. It can be observed that, the increase in $\delta n_{MAX}$ w.r.t $\alpha$ saturates at higher values of $\alpha$. (b) Variations of $|\delta n_{MAX}|$ as $\beta$ varies at $\alpha = 0.8$. (Details are given in SI)

These interactions span 6 orders of magnitude of the dissociation constant, ranging from μM for CaM-Ca$^{2+}$ and CaM-TFP to nM for CaM-M13 and pM for anti-IL2-IL2. The parameters used for calculation and the experimental data were taken from published BSI data[6b]. The complete list of parameters used in the simulation including the PDB IDs is presented in the SI text.

Figure 3 shows the comparison of simulated curves and experimental data of RI changes for the 4 interacting pairs extracted from[6b,]. In order to compare the temporal profile, we picked the largest target concentration used in the experiment for each of the protein pairs considered. The experimental calculation provides the change in RI directly. However, the experiment only reports the optical phase change. Therefore, we normalized the theoretical and experimental data with the saturated (equilibrium) values in order to compare the experimental and theoretical temporal profiles properly. As we see from Fig. 3, the calculated temporal profiles match experiments quite well except for the case of IL2-antibody interaction. Further, the optical phase change measured in the experiment should be linearly related to the RI change from basic textbook optics linking phase change to refractive index. To check this we compared the theoretically obtained RI change and the experimental phase change at saturation for the various target concentrations used in each of the protein interactions. From the inset figures in Fig. 3, we can clearly see this linearity between the experimental and simulated data. The close agreement between the temporal profiles as well as linear relationship between theoretical and experimental data provides a strong validation for our theoretical model. However, while the theoretical model captures the overall trends in the experiments very well, there is a large discrepancy when we consider the magnitude of the signal change, namely, the RI change corresponding to all the interactions except CaM-Ca$^{2+}$ is well below the detection limit of BSI which was reported to be around $10^{-6}$ RIU. Particularly striking is the case of IL2 interaction with its antibody. The predicted RI change for IL2-anti-IL2 interaction (about $10^{-9}$ RIU as seen in the x-axis of inset Fig. 3d) is 3 orders of magnitude lower than the experimental limit of detection which should have made this interaction undetectable in the BSI experiments. We emphasize that the discrepancy between the theoretical and experimental refractive index changes is not due to the uncertainty in the values chosen for $\alpha$ and $\beta$ factors in Eq. (15), which are the only free parameters in our model. The value of $\beta$ can be obtained with reasonable accuracy from published databases[10] and the effect of $\alpha$ on RI change is a monotonic increase with an upper bound, as evident from Eq. 16 (details in SI). Protein interactions span several orders of magnitude in affinities and consequently in the concentrations of the interacting species used in the experiments. From Eqs. (10) and (15), it is clear that the refractive index change depends upon these parameters. However, as we show, by scaling the RI change and the target concentration in a suitable manner, it is possible to separate the effect of variation in protein concentrations and $K_D$ from the effect of the free parameters related to RI change. The scaled variables are,

$$\delta n^{(SC)} = \frac{\delta n}{R_0} \quad (18a)$$

$$T_0^{(SC)} = \frac{T_0}{K_D} \quad (18b)$$

$$R_0^{(SC)} = \frac{R_0}{K_D} \quad (18c)$$

In Figure 4, we have plotted the maximum refractive index change (achieved at reaction saturation at equilibrium) against the scaled target concentration for various choices of the free parameters. This plot reaffirms the fact that the discrepancy between experimental and theoretical RI change values are not due to uncertainty in the choice of free parameters of the model because even the upper bound of refractive index change achieved at unrealistically large values of the free parameters is well below the reported detection limit of BSI.

It is evident that the maximum RI change depends on the receptor concentration used in the experiment. This dependence of RI change on receptor concentration is the reason why CaM-Ca$^{2+}$ interaction signal is about 3 orders of magnitude higher compared to anti-IL2-IL2 interaction signal. The receptor concentration ($R_0$) used in CaM-Ca$^{2+}$ and CaM-TFP interaction is of the order of a micro-molar whereas it is of the order of a nano-molar in the case of anti-IL2-IL2 leading to a 3 order of magnitude difference. The rescaling described in Eq. (18) eliminates the effect of concentration differences used in different protein interactions and focuses on the intrinsic differences in dielectric and conformational response between different interacting protein pairs leading to optical RI change as shown in Figure 5. Figure 5 shows that, for a given choice of α and β, in spite of a 6 order of magnitude difference in dissociation constants and corresponding receptor concentrations, rescaling according to Eq. (18) collapses all curves to a single curve (See SI more details). However, this data collapse happens only when the scaled receptor concentration is less than one, which is also a necessary condition for precise determination of dissociation constant as reported previously[17]. When the receptor concentration deviates from this range, such as the IL2-anti-IL2 experiment in [6b], the curves no longer collapse into a single one but still saturate to the same value as shown in Fig. 5(b). The value at saturation depends only on the choice of α and β and this value therefore will be different for different protein interaction pairs. The comparison between the scaled signal and concentration calculated for the protein pairs considered in previous experiments is shown in Fig. 5(c). The difference in saturation values reveal that while the CaM-Ca$^{2+}$, CaM-TFP and CaM-M13 interactions have similar optical response, the IL2-anti-IL2 interaction has a roughly 5 times higher optical response in comparison. However, this higher optical response is not sufficient to bridge the 3 orders of magnitude gap between the predicted RI change for the IL2-antibody interaction and the experimental limit of detection of the BSI technique. Finally, Fig. 5 (d) shows that the RI change is more sensitive to the α parameter compared to β but as already pointed out before, even an unphysically large value of α = 1000, cannot bridge the orders of magnitude gap between theory and experimental RI change.

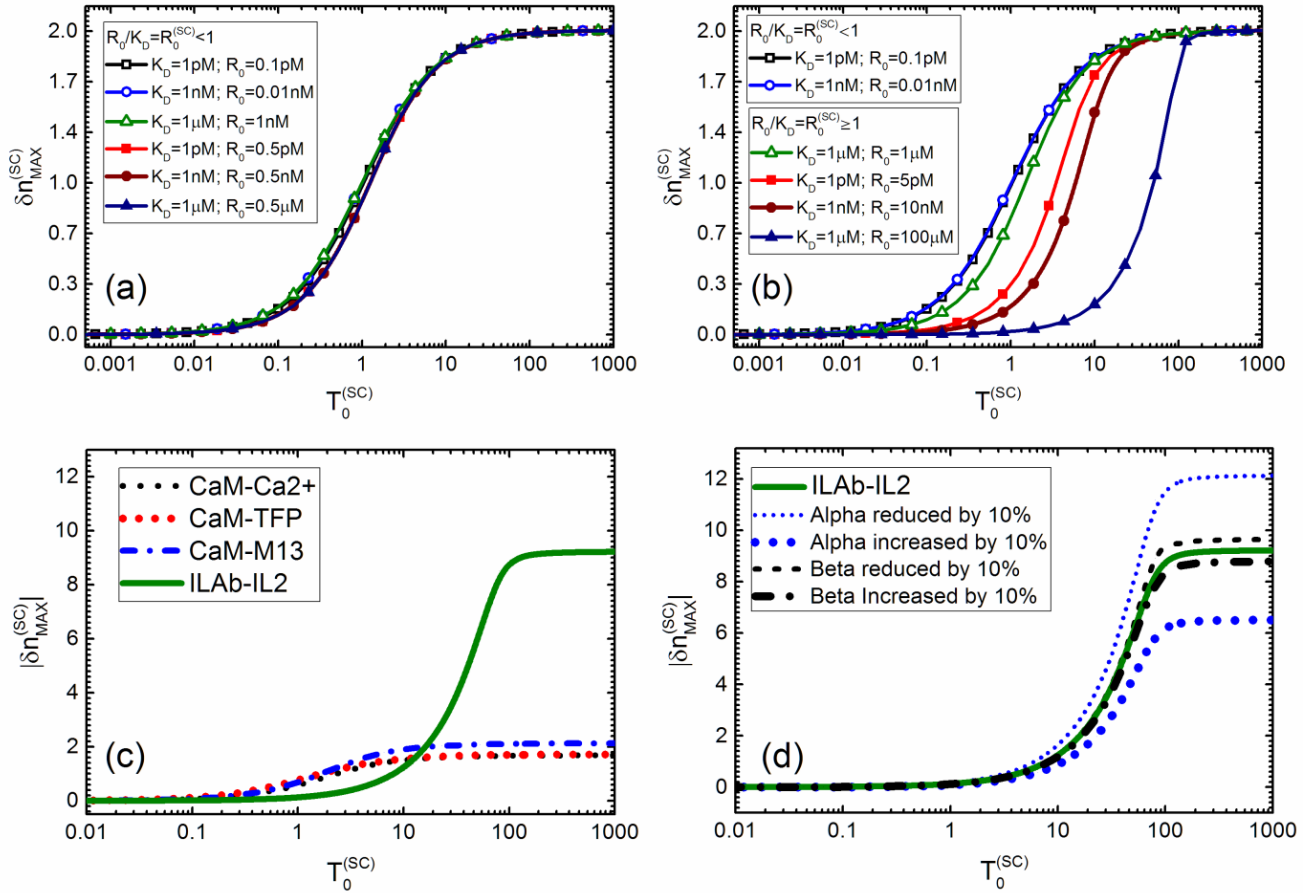

Figure 5. (a) Rescaling of variables according to Eq. (18) separates the effect of concentration variation from dielectric response in the experiments. Data across concentrations and affinities spanning orders of magnitude collapse to a single curve determined by the dielectric response. (b) This collapse is only possible when $R_0 < K_D$. Even when this condition is violated the saturated signal change is same (c) The scaled values computed for experimental data in[6b] and (d) RI change is more sensitive to alpha than beta.

## CONCLUSION AND DISCUSSION

We presented a first principles effective medium model to estimate the magnitude of refractive index changes arising due to protein interactions in free solution. We showed that by suitably rescaling the variables the effect of experimental parameters such as concentrations and affinities of the protein-pairs can be separated from the intrinsic properties responsible for the measured optical response. The implication of our study is that for interactions with low dissociation constants (in the nM – pM range) in free solution, the expected RI change is quite low $10^{-9}$ to $10^{-7}$ RIU which is below the reported detection limit of most optical sensing techniques described in literature including BSI. Hence, to explain the detection of specific protein interactions using BSI, one may have to look at alternate mechanisms such as local concentration enhancements or potential non-linearities from coupled effects. The analysis presented here provides a baseline requirement in terms of detection limit for systems that attempt to measure protein interactions in bulk solution. The analysis also makes clear why surface immobilization based methods have been so successful for molecular interaction measurements. This is because surface immobilization results in a large effective concentration of the molecular species due to immobilization in a small confined volume of a molecular monolayer. This effectively increases the value of $c_{RT}$ in Eq. (12) by 2-3 orders of magnitude causing the RI changes in surface immobilized methods to be higher than bulk values by a similar amount. This results in the surface immobilized RI changes to be of the order of $10^{-6}$–$10^{-4}$ RIU which is within the detection limit of current optical detection methods. Finally, we comment that temperature stability of the fluidic chamber may ultimately determine the smallest RI change detectable. The thermo-optic coefficient of DI water is around $10^{-5}$ RIU/C. The typical temperature stability of fluidic stability is of the order of 0.01 C, which sets a rough limit of $10^{-7}$ RIU for RI change. Given this practical limitation, optical confinement, for instance using plasmonic structures, to reduce the effective sampled volume (consequently increasing $c_{RT}$)[18] and/or molecular confinement using surface immobilization appears essential to push the limits of label-free optical sensing to detect weak interactions.

## ASSOCIATED CONTENT

**Supporting Information**

Derivations of equations; parameters used in the simulations. (PDF)

## AUTHOR INFORMATION

**Corresponding Author**

* E-mail: mvarma@iisc.ac.in

## NOTES

The authors declare no competing financial interest.

# Supporting Information

# Detection Limit for Optically Sensing Specific Protein Interactions in Free-Solution

Harish Sasikumar[†], Manoj M. Varma[*,†,‡,§]

[†]Department of Electrical Communication Engineering, Indian Institute of Science, Bangalore, 560012, India
[‡] Center for Nano Science and Engineering, Indian Institute of Science, Bangalore, 560012, India
[§] Robert Bosch Center for Cyber Physical Systems, Indian Institute of Science, Bangalore, 560012, India

## Table of Contents

# Equations

Symbolic Math Toolbox[1] was extensively used to solve the equations.

## Change in concentration due receptor-target interactions

The rate of increase in concentration of product ($\frac{dc_{RT}(t)}{dt}$) is aided by the current concentration of reactants ($c_R$ and $c_T$) and is hindered by the current concentration of product ($c_{RT}$)

$$\frac{dc_{RT}(t)}{dt} = k_{on}R(t)T(t) - k_{off}c_{RT}(t) \tag{1}$$

Concentration of a reactant at any time is the difference of initial reactant concentration and concentration of product.

$$R(t) = R_0 - c_{RT}(t) \tag{2a}$$
$$T(t) = T_0 - c_{RT}(t) \tag{2b}$$

Hence

$$\frac{dc_{RT}}{dt} = k_{on}(R_0 - c_{RT})(T_0 - c_{RT}) - k_{off}c_{RT} \tag{3}$$

$$\frac{dc_{RT}}{dt} = k_{on}c_{RT}^2 - (k_{on}(R_0 + T_0) + k_{off})c_{RT} + k_{on}R_0T_0 \tag{4}$$

The above equation can be solved with initial condition $c_{RT}(0) = 0$ to obtain

$$c_{RT}(t) = \frac{R_0 + T_0 + K_D}{2} - \frac{D}{2}\tanh\left(\frac{k_{on}D}{2}t + \tanh^{-1}\left(\frac{R_0 + T_0 + K_D}{D}\right)\right) \tag{5}$$

Where

$$D = \sqrt{(R_0 - T_0)^2 + 2K_D(R_0 + T_0) + K_D^2} \tag{6}$$

$K_D$, referred to as the dissociation constant of the reaction, is the ratio of the off-rate to on-rate

$$K_D = \frac{k_{off}}{k_{on}} \tag{7}$$

*Equilibrium concentration of receptor-target complex*

We have the relation

$$\lim_{t \to \infty} \tanh(A\,t + B) = 1;\, A > 0 \tag{8}$$

Hence

$$C_{max} = \lim_{t \to \infty} c_{RT}(t) = \frac{R_0 + T_0 + K_D - D}{2} \tag{9}$$

## Special Case: High target concentration

In case of high target concentration or when the target concentration is kept constant, $T(t) \approx T_0$ and Eqn. (1) reduces to,

$$\frac{dc'_{RT}}{dt} = k_{on}R_0 T_0 - (k_{on}T_0 + k_{off})c'_{RT} \tag{10}$$

The above equation can be solved with initial condition $c_{RT}(0) = 0$ to obtain

$$c'_{RT}(t) = \frac{k_{on}R_0 T_0}{k_{off} + k_{on}T_0}\left(1 - \exp(-t(k_{off} + k_{on}T_0))\right) \tag{11}$$

*Equilibrium concentration of receptor-target complex*

We have the relation

$$\lim_{t \to \infty} e^{(-A\,t + B)} = 0;\, A > 0 \tag{12}$$

Hence

$$C'_{max} = \lim_{t \to \infty} c_{RT\_High}(t) = \frac{k_{on}R_0 T_0}{k_{off} + k_{on}T_0} \tag{13}$$

## Volume fraction

Volume fraction,

$$f_R \equiv \frac{V_R}{V}$$

Where, $V_R$ is the total volume occupied by receptor molecules and $V$ is the total volume of the solution.

$$V_R = Nv_R$$

where, $N$ is the total number of receptor molecules and $v_R$ is the volume of a receptor molecule. Hence

$$f_R = \frac{N v_R}{V}$$

Total number of receptor molecule,

$$N = N_M N_{AVO}$$

where $N_M$ is the total number of moles of receptor and $N_{AVO}$ is the Avogadro's Number,

$$f_R = \frac{N_M N_{AVO}}{V} v_R$$

Molar concentration, $C_R \equiv \frac{N_M}{V}$. Hence,

$$f_R = c_R N_{AVO} v_R \tag{14}$$

## Refractive index of proteins

The concentration of receptors, $c_R$, can be written in terms of the mass density of receptors in the solution, $\mu_R$ and the molecular mass $M_R$ as $c_R = \mu_R/M_R$ which allows us to rewrite Eq. (14) as,

$$f_R = \frac{\mu_R}{\rho_R} \tag{15}$$

where $\rho_R$ is the density of the receptor molecule.

We can define a quantity "refractive index increment", $\gamma = \frac{\partial n_{eff}}{\partial \mu_R}$ which, along with Eq. (15) and the following expression for effective refractive index of the medium,

$$n_{eff} = n_m \left(1 + \frac{3}{2} f_R \left(\frac{\varepsilon_R - \varepsilon_m}{\varepsilon_R + 2\varepsilon_m}\right)\right) \tag{16}$$

can be written as,

$$\gamma = \frac{3}{2} \frac{n_m}{\rho_R} \left(\frac{\varepsilon_R - \varepsilon_m}{\varepsilon_R + 2\varepsilon_m}\right) \tag{17}$$

The quantity $\gamma$ has been measured for several proteins and is found to be tightly clustered around the value of about 0.19 ml/g[2]. The universality of quantity, in the light of Eq. (17), must lie in the tight clustering of the density and dielectric permittivity of proteins. Indeed the density of proteins is tightly clustered around 1.35 gm/cc[3]. Using these numbers in Eq. (17) yields the relative dielectric permittivity of proteins to be tightly clustered around 2.6, or in terms of refractive index, around 1.6. In other words we can consider any protein solution to be composed of dielectric spheres with relative permittivity of about 2.6 and molecular volume corresponding to the protein molecule being considered.

## Change in refractive index due to small changes in volume fraction of solute

The effective permittivity $\epsilon_{eff}$ of the solution, based on Maxwell Garnett effective medium theory is given by,

$$\frac{\varepsilon_{eff} - \varepsilon_m}{\varepsilon_{eff} + 2\varepsilon_m} = f \frac{\varepsilon_R - \varepsilon_m}{\varepsilon_R + 2\varepsilon_m} \tag{18}$$

$$n_{eff} = \sqrt{\varepsilon_{eff}} = \sqrt{\varepsilon_m \frac{\varepsilon_R(1 + 2f_R) + 2\varepsilon_m(1 - f_R)}{\varepsilon_R(1 - f_R) + \varepsilon_m(2 + f_R)}} \tag{19}$$

$$n_{eff} \approx n_m \left(1 + \frac{3}{2} f_R \left(\frac{\varepsilon_R - \varepsilon_m}{\varepsilon_R + 2\varepsilon_m}\right)\right) + O(f_R^2) \tag{20}$$

Change in refractive index,

$$\delta n = n_{eff} - n_m = \frac{3}{2} n_m f_R \left(\frac{\varepsilon_R - \varepsilon_m}{\varepsilon_R + 2\varepsilon_m}\right) \tag{21}$$

## Equilibrium Refractive index change

$$\delta n_{MAX} = FC_{max}(v_R + v_T)\Delta \quad (22)$$

Where, F is a constant for a medium, $F = 1.5\, n_m N_{Avo} = 1.204 \times 10^{24}$, $n_m$ is the refractive index of the medium (water), taken as 1.333 and $N_{Avo}$ is the Avogadro's number ($6.022 \times 10^{23}$). $\Delta$ is the factor capturing the effect of optical and volumetric changes in the molecules.

$$\Delta = \left(\beta \frac{\alpha\varepsilon_P - \varepsilon_m}{\alpha\varepsilon_P + 2\varepsilon_m} - \frac{\varepsilon_P - \varepsilon_m}{\varepsilon_P + 2\varepsilon_m}\right) \quad (23)$$

$C_{max}$ is the equilibrium value of concentration as given in Eqn. (9)

$$\delta n_{MAX} = \frac{R_0 + T_0 + K_D - \sqrt{(R_0 - T_0)^2 + 2K_D(R_0 + T_0) + K_D^2}}{2}[F(v_R + v_T)\Delta] \quad (24)$$

In case of high target concentration, Eqn. (22) can be written by substituting Eqn. (13)

$$\delta n'_{MAX} = \frac{k_{on} R_0 T_0}{k_{off} + k_{on} T_0}[F(v_R + v_T)\Delta] \quad (25)$$

## Scaling

From Eqn. (22), Eqn. (9) and Eqn. (6), $\delta n_{MAX}$ can be written as

Defining

$$\delta n_{MAX}^{(SC)} = \frac{\delta n}{R_0} \quad (26a)$$

$$T_0^{(SC)} = \frac{T_0}{K_D} \quad (26b)$$

$$R_0^{(SC)} = \frac{R_0}{K_D} \quad (26c)$$

Makes Eqn. (24) as

$$\delta n_{MAX}^{(SC)} = \frac{1 + \frac{1 + T_0^{(SC)}}{R_0^{(SC)}} - \sqrt{\left(1 - \frac{T_0^{(SC)}}{R_0^{(SC)}}\right)^2 + \frac{2}{R_0^{(SC)}}\left(1 + \frac{T_0^{(SC)}}{R_0^{(SC)}}\right) + \left(\frac{1}{R_0^{(SC)}}\right)^2}}{2}[F(v_R + v_T)\Delta]K_D \quad (27)$$

In case of high target concentration, $T_0 \gg R_0$ or equivalently $T_0^{(SC)} \gg R_0^{(SC)}$, the expression becomes

$$\delta n'^{(SC)}_{MAX}\bigg|_{T_0^{(SC)} \gg R_0^{(SC)}} = \frac{T_0^{(SC)} K_D}{T_0^{(SC)} + 1}[F(v_R + v_T)\Delta] \quad (28)$$

# Simulation Parameters

## Constants

**Table S- 1: Constants and their values used for the simulations.**

| Constants | Explanation | Value |
|---|---|---|
| $n_m$ | Refractive index of medium (water) | 1.333 |
| $\varepsilon_m$ | Dielectric permttivity of medium (water) | 1.777 |
| $\varepsilon_p$ | Dielectric permttivity of protein | 2.6 |
| $N_{Avo}$ | Avogadro's Number | $6.022 \times 10^{23}$ |
| F | Constant part in the expression for refractive index change δn.<br>$F = 1.5\, n_m N_{Avo}$ | $1.204 \times 10^{24}\ M^{-1}$ |

## Rate & Equilibrium Constants

The rate constants are extracted from the supporting online material provided with [4], using the procedure given in the same material. Data extraction from the images were done using Grabit[5], a code obtained from File Exchange (MATLAB Central).

**Table S- 2. Rate constants ($k_{on}$, $k_{off}$ and $K_D$) extracted from the kinetic trace[4]**

| Interacting Proteins | Calmodulin (CaM) & | | | interleukin2 antibody (IL2-Ab) |
|---|---|---|---|---|
| | $Ca^{2+}$ | Trifluoperazine Dihydrochloride (TFP) | M13 peptide | IL-2 |
| Association rate constant ($k_{on}$) | $7.5 \times 10^3$ M⁻¹s⁻¹ | $1.8 \times 10^4$ M⁻¹s⁻¹ | $3.1 \times 10^7$ M⁻¹s⁻¹ | $1.2 \times 10^9$ M⁻¹s⁻¹ |
| Dissociation rate constant ($k_{off}$) | $2.7 \times 10^{-2}$ s⁻¹ | $8.3 \times 10^{-2}$ s⁻¹ | $9.4 \times 10^{-2}$ s⁻¹ | $3.2 \times 10^{-2}$ s⁻¹ |
| Equilibrium dissociation constant ($K_D$) | $3.6 \times 10^{-6}$ M | $4.7 \times 10^{-6}$ M | $3.1 \times 10^{-9}$ M | $2.5 \times 10^{-11}$ M |

## Estimating Molecular Volume

Estimates for volumes of the receptors and receptor-ligand complexes were obtained using two approaches. In the first approach, values of radius of gyration ($R_g$) were obtained from the supporting information provided with[6]. The volumes were approximated to using the relation $V_{R_g} = \frac{4}{3}\pi R_g^3$. The second approach was based on[7], where the volume of the protein molecule was estimated from molecular weight ($M_w$) as $V_{M_w} = 1.22 \times 10^{-27} \left(\frac{L}{Da}\right) \times M_w$ (Da). Molecular weights were obtained from the protein data bank[8] for the PDB IDs given in supporting information of[6]. These estimates are comparable to the actual values used in the simulation. The values used in the simulation, $\upsilon_R + \upsilon_T$ and $\upsilon_{RT}$ were obtained using chimera[9].

**Table S-3. Estimating molecular volumes of receptors and receptor-ligand complexes.**

| | Receptor | Calmodulin (CaM) | | | IL2-Ab |
|---|---|---|---|---|---|
| | Ligand | Ca$^{2+}$ | TFP | M13 peptide | IL2 |
| Unbound | PDB ID | 1CFD | 1OSA | 1OSA | 1M4C |
| | Radius of gyration (Å) | 20.29 | 22.45 | 22.45 | 38.98 |
| | Molecular weight (Da) | 16721.46 | 16847.86 | 16847.86 | 30872.21 |
| | Est. Mol. Vol (from Rg) [nm³] | 35.0 | 47.4 | 47.4 | 248 |
| | Est. Mol. Vol (from Mw) [nm³] | 20.2 | 20.4 | 20.4 | 37.4 |
| | $\upsilon_R + \upsilon_T$ [L] | $1.74 \times 10^{-23}$ | $1.74 \times 10^{-23}$ | $2.11 \times 10^{-23}$ | $8.41 \times 10^{-23}$ |
| Bound | PDB ID | 1OSA | 1LIN | 1CDL | 4YUE |
| | Radius of gyration (Å) | 22.45 | 15.54 | 16.50 | 28.11 |
| | Molecular weight (Da) | 16847.86 | 18511.77 | 76153.21 | 65570.34 |
| | Est. Mol. Vol (from Rg) [nm³] | 47.4 | 15.7 | 18.8 | 93.0 |
| | Est. Mol. Vol (from Mw) [nm³] | 20.4 | 22.4 | 92.1 | 79.3 |
| | $\upsilon_{RT}$ [L] | $1.73 \times 10^{-23}$ | $1.66 \times 10^{-23}$ | $1.95 \times 10^{-23}$ | $6.67 \times 10^{-23}$ |
| | $\beta = \upsilon_{RT}/(\upsilon_R + \upsilon_T)$ | 0.994 | 0.954 | 0.924 | 0.7931 |

# Figures

## Figure 2: Variation of Δ

Plot of Eq. (23)

$$\Delta = \left(\beta \frac{\alpha \varepsilon_P - \varepsilon_m}{\alpha \varepsilon_P + 2\varepsilon_m} - \frac{\varepsilon_P - \varepsilon_m}{\varepsilon_P + 2\varepsilon_m}\right)$$

With values of $\varepsilon_P$ and $\varepsilon_m$ taken from Table S- 1.

From the equation, we can see that the fraction involving α is

$$T = \frac{\alpha \varepsilon_P - \varepsilon_m}{\alpha \varepsilon_P + 2\varepsilon_m}$$

It becomes finite at both the extremes of α.

$$\lim_{\alpha \to 0} T = -0.5$$

$$\lim_{\alpha \to \infty} T = 1$$

# Figure 3 : Temporal evolution of RI change

The simulation plots are according to the equation

$$\delta n = F\left(\frac{R_0 + T_0 + K_D}{2} - \frac{D}{2}\tanh\left(\frac{k_{on}D}{2}t + \tanh^{-1}\left(\frac{R_0 + T_0 + K_D}{D}\right)\right)\right)(v_R + v_T)\left(\beta\frac{\alpha\varepsilon_P - \varepsilon_m}{\alpha\varepsilon_P + 2\varepsilon_m} - \frac{\varepsilon_P - \varepsilon_m}{\varepsilon_P + 2\varepsilon_m}\right) \quad (29)$$

Value of the constant, F is used as given in Table S- 1. Rate constants ($k_{on}$, $k_{off}$ and $K_D$) are taken from Table S- 2. Values of $v_{RT}$ and $\beta$ are taken from Table S- 3. As reasonable value, $\alpha$ is taken as 0.8.

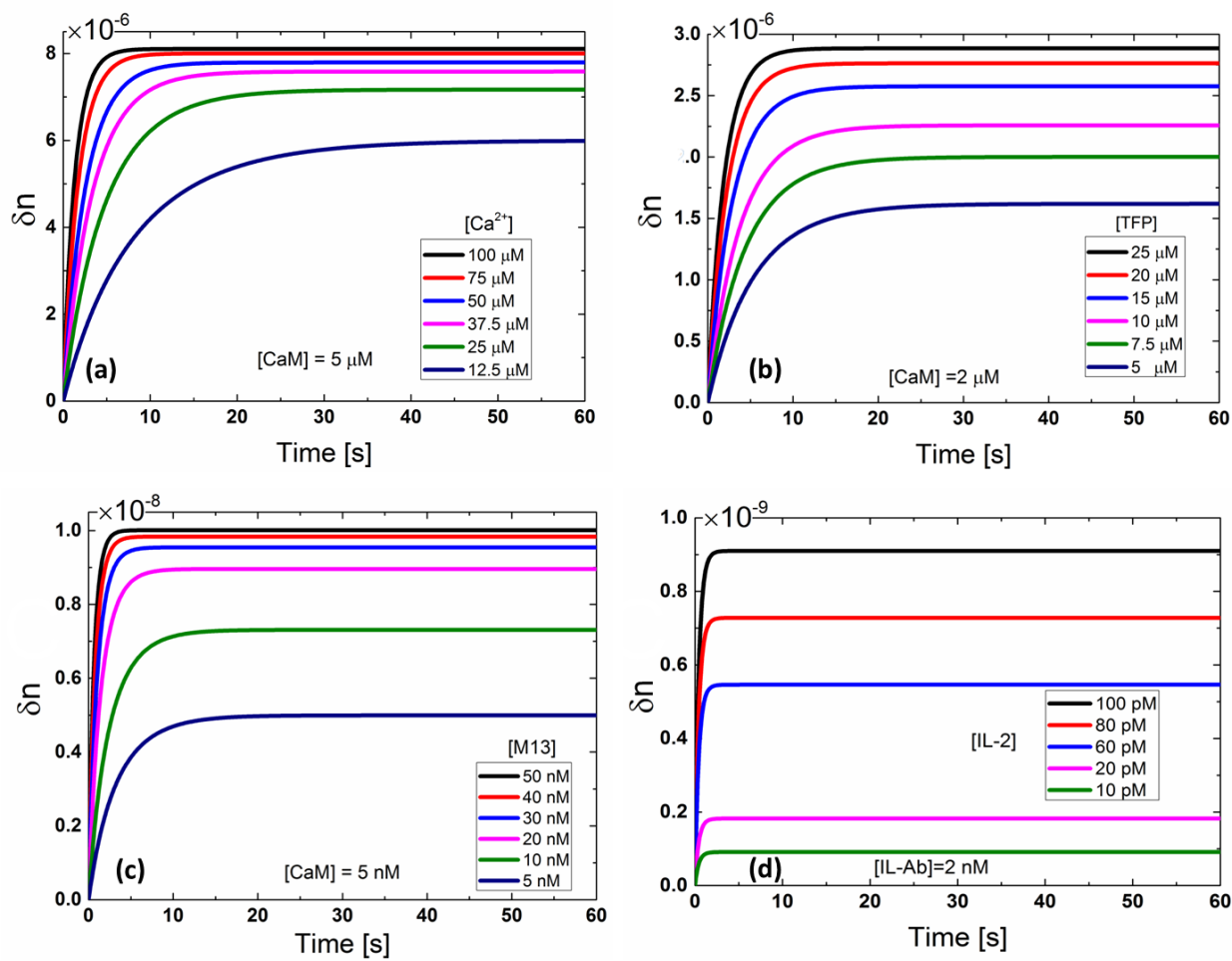

**Figure S- 1. RI changes computed for experiments in[4]**

In the main text, the simulated curves were compared with the experimental curves (of refractive index changes) reported by[4], after normalizing with the maximum values.

## Figure 4 (a): Effect of Alpha

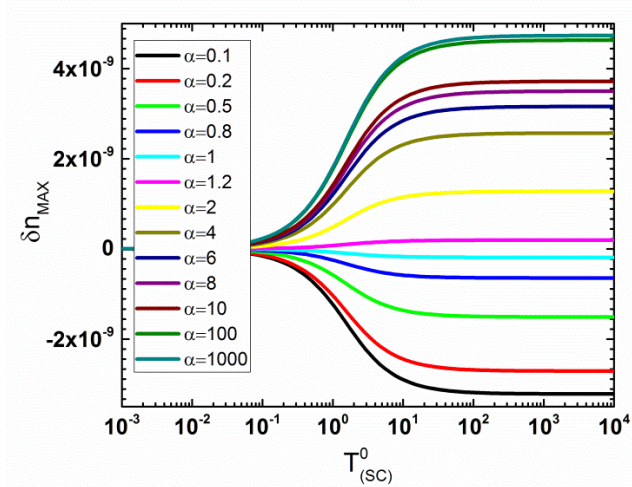

**Figure S- 2: Effect of α**

Value of the constants, $F, N_{Avo}, n_m, \varepsilon_m, \varepsilon_p$ are used as given in Table S- 1. Values of $v_R + v_T$ is taken as $1.7 \times 10^{-23}$ L (from Table S- 3). As reasonable value, $\beta$ is taken as 0.8. $R_0$ and $K_D$ are taken as 2 nM. As shown in Figure S- 2, $\delta n_{MAX}$ reaches 0 at $1 < \alpha < 1.2$. Hence $|\delta n_{MAX}|$ reaches a minimum at this interval.

## Figure 4 (b): Effect of Beta

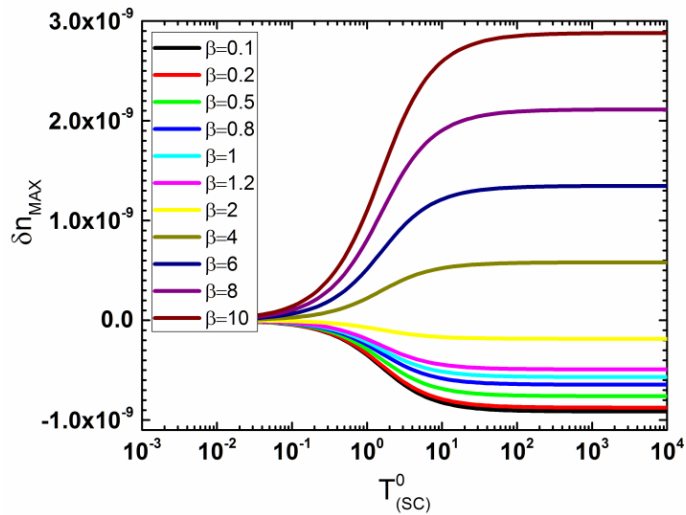

**Figure S- 3: Effect of β**

Value of the constants, $F, N_{Avo}, n_m, \varepsilon_m, \varepsilon_p$ are used as given in Table S- 1. Values of $v_R + v_T$ is taken as $1.7 \times 10^{-23}$ L (from Table S- 3). As reasonable value, $\alpha$ is taken as 0.8. $R_0$ and $K_D$ are taken as 2 nM. As shown in Figure S- 3, $\delta n_{MAX}$ reaches 0 at $2 < \beta < 4$. Hence $|\delta n_{MAX}|$ reaches a minimum at this interval.

## Figure 5(a,b): Collapse of scaled refractive index change curves

The plot is according to Eqn. (24) and Eqn. (26). Value of $(v_R + v_T)\Delta$ is taken to be $1.7 \times 10^{-24}$ L.

## Figure 5(c): Plot of scaled refractive index change at experimental conditions

The plot is according to Eqn. (27). Value of the constant, F is used as given in Table S- 1. As reasonable value, α is taken as 0.8. Values of $(v_R + v_T)$ and β, for the calculation of Δ, are taken from Table S- 3.

From α and β, Δ is calculated from Eq. (23). Correspondingly, the value of Δ can be obtained from Figure 2 in the main text.

Table S- 4. Values of parameters used in the figure of normalized plots. (From [4])

| Receptor-Ligand | Calmodulin (CaM)- $Ca^{2+}$ | Calmodulin (CaM)- TFP | Calmodulin (CaM)- M13 peptide | IL2-Ab – IL2 |
|---|---|---|---|---|
| $R_0$ | 5 μM | 2 μM | 5 nM | 2 nM |
| $K_D$ | 3.6 μM | 4.61 μM | 3.03 nM | 26.7 pM |
| $R_0^{(SC)}$ | 1.389 | 0.434 | 1.650 | 74.906 |
| $(v_R + v_T)\Delta$ [L] | $1.40 \times 10^{-24}$ | $1.43 \times 10^{-24}$ | $1.77 \times 10^{-24}$ | $7.66 \times 10^{-24}$ |

## Figure 5(d): Effect of 10% variations

The plot is according to Eqn. (27). Value of the constant, F is used as given in Table S- 1. As reasonable value, **α** is taken as 0.8. Values of $(v_T + v_T)$ and **β**, for the calculation of **Δ**, are taken from Table S- 3.

Table S- 5. Values of parameters used in the figure of normalized plots for IL2-Ab

| IL2-Ab – IL2 | At calculated β(0.7931) and α = 0.8 | 10% reduced α (0.72) | 10% increased α (0.88) | 10% reduced β (0.714) | 10% increased β (0.714) |
|---|---|---|---|---|---|
| $\frac{(v_R + v_T)\Delta}{7.66 \times 10^{-24}}$ [L] | 1 | 1.3160 | 0.7066 | 1.0471 | 0.9535 |